\documentclass[journal,twocolumn]{IEEEtran}

\makeatletter
\usepackage{epsfig, psfrag}
\usepackage{subfigure}
\usepackage{epstopdf}
\usepackage{psfrag}
\usepackage{latexsym}
\usepackage{amssymb}
\usepackage{amsthm} 
\usepackage{graphics,color}
\usepackage{graphicx}
\usepackage{amsmath}
\usepackage{cite}
\usepackage{url}
\usepackage{xspace}
\usepackage[bookmarks]{hyperref}
\usepackage{enumerate}

\newtheorem{theorem}{Theorem}
\newtheorem{lemma}{Lemma}

\newtheorem{corollary}{Corollary}

\theoremstyle{definition}
\newtheorem{definition} {Definition}
\newtheorem{remarks}{Remark}

\newcommand{\p}{\bar{p}}
\def\P{\mathsf{P}}
\def\expe{\mathbb{E}}   

\def\ber{\mathsf{Ber}}

\providecommand{\secref}[1]{Sec.~\ref{#1}}

\providecommand{\figref}[1]{Fig.~\ref{#1}}
\providecommand{\colref}[1]{Corol.~\ref{#1}}

\def\mc{\mathcal}

\def\mbb{\mathbb}

\usepackage{mathtools}
\mathtoolsset{showonlyrefs=true}

\usepackage{ifthen}
\newcommand{\VersionLength}{long}

\providecommand{\ver}{\ifthenelse{\equal{\VersionLength}{long}}}
\providecommand{\nver}{\ifthenelse{\equal{\VersionLength}{short}}}

\usepackage{tikz}
\usetikzlibrary{shapes,arrows,decorations.markings,positioning}

\tikzstyle{plant} = [draw, fill=red!5, rectangle, 
    minimum height=3em] 
\tikzstyle{block} = [draw, fill=blue!5, rectangle, 
    minimum height=3em] 
\tikzstyle{sum} = [draw, fill=yellow!10, circle, node distance=1cm]
\tikzstyle{coord} = [coordinate]
\tikzstyle{gain} = [draw, fill=red!5, regular polygon, regular polygon sides=3, shape border rotate=-90]
\tikzstyle{pinstyle} = [pin edge={to-,thick,black}]

\IEEEoverridecommandlockouts

\begin{document}

\allowdisplaybreaks

\title{Causal Posterior Matching and its Applications}

\ver{
\author{Anusha Lalitha, Anatoly Khina, and Tara Javidi
    \thanks{A.~Lalitha is with the Department of Electrical Engineering, Stanford University, Palto Alto, CA~94305, USA (e-mail: \texttt{alalitha@stanford.edu}).}
    \thanks{T.~Javidi is with the Department of Electrical and Computer Engineering, University of California, San Diego, La Jolla, CA~92093, USA (e-mail: \texttt{tjavidi@eng.ucsd.edu}).}
	\thanks{A.~Khina is with the School of Electrical Engineering, Tel Aviv University, Tel Aviv, Israel~6997801 (e-mail: \texttt{anatolyk@eng.tau.ac.il}).}
} 
}{
\author{
    \IEEEauthorblockN{Anusha Lalitha}
    \IEEEauthorblockA{Electrical \& Computer Engineering \\
                  University of California, San Diego \\
                  La Jolla, CA~92093, USA \\
                  Email: {\em alalitha@eng.ucsd.edu}
                 }
    \and
    \IEEEauthorblockN{Anatoly Khina, Victoria Kostina}
    \IEEEauthorblockA{Electrical Engineering \\
					California Institute of Technology \\
                  Pasadena, CA~91125, USA \\
                  Email: {\em \{khina,vkostina\}@caltech.edu}
                 }
    \and
    \IEEEauthorblockN{Tara Javidi}
    \IEEEauthorblockA{Electrical \& Computer Engineering \\
                  University of California, San Diego \\
                  La Jolla, CA~92093, USA \\
                  Email: {\em tjavidi@eng.ucsd.edu}
                 }
} 
}

\newcommand{\khina}{A.~Khina}
\newcommand{\invited}{}
\newcommand{\PI}{}
\newcommand{\Student}{}
\newcommand{\CoRes}{}


\maketitle
\begin{abstract}
We consider the problem of communication over the binary symmetric channel with feedback, where the information sequence is made available in a causal, possibly random, fashion. We develop a real-time variant of the renowned Horstein scheme and provide analytical guarantees for its error-probability exponential decay rate. We further use the scheme to stabilize an unstable control plant over a binary symmetric channel and compare the analytical guarantees with its empirical performance as well as with those of anytime-reliable codes.
\end{abstract}

\section{Introduction}
\label{sec:intro}


While feedback cannot increase the capacity of memoryless channels \cite[Ch.~7.12]{CoverBook2Edition}, 
it can dramatically reduce the probability of error and the complexity of the communication schemes that achieve them. 
For the binary symmetric channel (BSC),
a \textit{horizon-free} sequential scheme was proposed by Horstein~\cite{Horstein:BSC:IT1963};
it was rigorously proved to attain capacity by Shayevitz and Feder \cite{PosteriorMatching} for this and other channels, via its generalization---the \textit{posterior matching} (PM) scheme. 
Exponential error-probability guarantees, for the \textit{finite-horizon} setting, 
were constructed in \cite{BurnashevZigangirov1974,D'yachkov:PPI1975,Li-ElGamal:IT2015,NaghshvarJavidiWigger:IT2015}. 
An exponential bound on the error probability in the horizon-free case has been devised by Waeber et al.~\cite{WaeberFrazierHenderson:SICON2013}, 
although this bound becomes trivial for rates much below the capacity.


The availability of instantaneous noiseless feedback obviates the need of transmitting long error-correcting codes across long epochs,
and enables instead the use of sequential communication schemes, 
by providing full knowledge of the receiver's state to the transmitter. 
%
A class of problems where this may have powerful implications is that of stabilizing an unstable control plant over a noisy channel.
In particular, in the presence of feedback, the structure of the horizon-free PM decoder seems to match the structure of anytime reliable decoders (proposed for stabilizing unstable linear plants over noisy channel \cite{SahaiMitterPartI,SukhavasiHassibi,TreeCodes:ISIT2016}). 

However, the classical PM schemes assume that the entire  information (possibly infinite bit) sequence is available essentially non-causally to the transmitter, 
prior to the beginning of transmission. 
That is, they are sequential with respect to the transmitted sequence (codeword) but not with respect to the information sequence.
Consequently, the non-causal knowledge assumption precludes the use of the classical PM scheme for real-time and control scenarios, in which the data to be transmitted is determined in a causal fashion.

In the current work, we consider a real-time setting, described in detail in Section~\ref{sec:model}, in which the bits arrive to the transmitter one-by-one. In this section, we first consider the case where the inter bit-arrival rate is deterministic. We construct, in Section~\ref{sec:PM}, a causal (horizon-free) PM scheme for this setting, i.e., a scheme that is sequential with respect to both the information and the transmitted sequences. We provide exponential guarantees for the error probability akin to those 
of~\cite{WaeberFrazierHenderson:SICON2013}, in Section~\ref{sec:main_res}. We apply the proposed scheme, in \secref{ss:NCS}, for control over a BSC with feedback
and compare its analytic and empirical stabilization performance with those of the anytime-reliable codes of Sahai and Mitter \cite{SahaiMitterPartI} that use no feedback but are computationally demanding, as well as with those of Simsek et al.~\cite{SimsekJainVaraiya:AC2004} in Section~\ref{sec:numeric}. Analytic guarantees for the scheme of \cite{SimsekJainVaraiya:AC2004} exist only for the case in which the entire information sequence is known in advance, which corresponds, to the case of stabilizing an unstable linear system with possibly unknown initial conditions but with no system disturbance. Next, we extend apply for causal PM strategy to the case where the information bits arrive at
random times, under the assumption that the inter-arrival times (time-arrival differences) have a known finite support. We conclude the paper with a discussion, in Section~\ref{sec:summary}.


\underline{\textit{Notation:}} 
$\mbb{N}$ denotes the set of natural numbers. For $k,t \in \mbb{N}$, $k < t$, the sequence $\{s_k, s_{k+1}, \ldots, s_t\}$  is denoted as $s_k^t$. For $M \in \mbb{N}$, the sequence of integers $\{1,2,\ldots, M\}$ is denoted $[M]$. 
The binary entropy of probability $p$ is denoted by $h(p) = -p \log p - \p \log \p$ with $\p := 1-p$; 
all logarithms in this work are to the base 2. For any probability mass function (pmf) $p$, let $p^{\bigotimes i}$ denote $p$ convolved with itself $i$ times.

\section{Problem Formulation}
\label{sec:model}

The transmitter wishes to transmit an infinite stream of bits over a BSC with cross over probability $p \in (0, {1}/{2})$. Let $s_i \in \{0,1\}$ denote the $i$-th bit in the infinite bit stream where we assume $s_i \sim \ber{(1/2)}$. In our notation, we think of the infinite bit sequence as the binary expansion of a single message point $\Theta$ uniformly distributed over the unit interval i.e., $\Theta \sim \text{Unif}[0,1)$. We assume that the bits of the message point $\Theta$ are revealed to the transmitter causally. A new bit is revealed after every fixed $n \in \mbb{N}$ time steps. In other words, the $i$-th bit arrives at time $T_i := n(i-1)+1$ for all $i \geq 2$ with $T_1 = 1$. For all time instants $t \in \mbb{N}$, define the following random variable
\begin{align}
    b(t):= \max\{i \in \mbb{N}: T_{i} \leq t\}.
\end{align}
In other words, $b(t)$ denotes the number of bits that have arrived by time $t$, and for fixed inter bit-arrival time of $n$ we have $b(t) = \left\lceil \frac{t}{n} \right\rceil$ for all $t$.

We now define the feedback communication scheme of a causally available information bit sequence, where a new bit is made available to the encoder after $n$ time steps. We assume that receiver knows the fixed arrival times of each bit in the infinite bit-stream, i.e., decoder has the knowledge of $n$. 

\begin{definition}[Causal encoder]
The causal encoder $\mc{E}$ is described by a sequence of (causal) functions  $\{\mc{E}_{t}\}_{t \geq 1}$. A causal encoder with feedback emits a channel input symbol $x_{t} \in \{0,1\}$ as a function of number of bits available $s_1^{b(t)}$ and past channel outputs $y^{t-1}_1$:
\begin{align}
    x_{t} = \mc{E}_{t}\left( s_1^{b(t)}, y^{t-1}_1\right).
\end{align}
\end{definition}

\begin{definition}[Causal decoder]
The decoder $\mc{D}$ is described by the sequence of functions $\{\mc{D}_{t}\}_{t \geq 1}$. After observing $t$ channel outputs,
the decoder outputs a vector of estimates of all the bits available at the encoder thus far, $\hat{s}^{b(t)}_1(t) = [\hat{s}_1(t), \hat{s}_2(t), \ldots, \hat{s}_{b(t)}(t)] \in \{0, 1\}^{b(t)}$:
\begin{align}
\hat{s}_1^{b(t)}(t) = \mc{D}_{t}\left(y^{t}_1\right).
\end{align}
\end{definition}

At any time instant $t$, we want to analyze the probability of error in decoding the first $j$ bits $ \P\left(\hat{s}_1^j(t) \neq s_1^j\right)$ for $1 \leq j \leq b(t)$. Since the bits that arrive early get encoded for longer duration it is natural to expect that the probability of error in decoding the older bits is smaller than that in decoding the newer bits.

\section{Causal Posterior Matching Strategy}
\label{sec:PM}

In this section, we propose a causal PM based encoding and decoding strategy to transmit a causally available message where the inter bit-arrival times are random.

First, we provide an overview of the strategy. At time $t$, suppose only the first $i$ bits are available to the encoder i.e., consider the event $b(t) = i$. Consider the unit interval $[0,1]$ and divide it into bins of equal length $2^{-i}$. The indices of the bins are enumerated by the elements of $\{0,1,\ldots, 2^{i}-1\}$. Since, the message point $\Theta$ is located on the unit interval, the first $i$ bits $s_1^{i}$ of the infinite bit-stream provide the index of the bin containing $\Theta$. The encoder and decoder maintain a posterior probability of the message point $\Theta$ belonging to each bin given the past channel outputs. Until a new bit arrives, we use causal posterior matching (described in detail in Sections~\ref{sec:enc} and \ref{sec:dec} below) to encode the first $i$ bits. After observing each channel output the decoder and encoder (using feedback) perform a Bayesian update to the posterior probability of $\Theta$. When a new bit arrives, we divide each bin from the previous $2^{i}$ bins into $2$ equal bins, resulting in $2^{i+1}$ bins in total. Furthermore, we divide the posterior probability equally into the newly created bins. Now, the first $i+1$ bits provide the index of the bin containing $\Theta$ on a grid with $2^{(i+1)}$ bins. This process of dividing the existing bins and the posterior probability to accommodate a new bit continues in a horizon-free manner. At any time $t$, the binary expansion of the index of the bin that contains the median of the posterior distribution are declared as estimates of the bits available at the encoder.

\subsection{Preliminaries}
Let BSC$(p)$ denote a BSC with cross-over probability $p \in (0, {1}/{2})$ with input $X \in \{0,1\}$, output $Y \in \{0,1\}$:
\begin{align}
    \P(Y = y|X = x) = 
    \begin{cases}
        p & \text{if } y \neq x,\\
        \p  & \text{if } y = x .
    \end{cases}
\end{align} 
Let $C(p) := 1-h(p)$ denote the capacity of BSC$(p)$. 

For all $t \geq n(i-1)+1$, the encoder has access to the first $i$ bits. Furthermore, the decoder maintains a posterior distribution of $\Theta$ after observing past $t$ channel outputs $y_1^t$, i.e., $\P_{\Theta|Y_1^{t}}\left(\Theta \in [(k-1)2^{-i}, k 2^{-i}) \middle| y_1^t \right)$ for all $k \in \{0, \ldots, 2^{i}-1\}$. Let $F_{\Theta|Y_1^{t}}$ denote the cumulative distribution function (CDF) of posterior probability distribution. Due to the presence of feedback, the posterior distribution maintained by the decoder is available to the encoder as well. We refer to the point $F^{-1}_{\Theta|Y_1^{t}} \left( {1}/{2} \middle| y_1^t \right) \in [0,1)$ as the median of the posterior probability distribution at time~$t$. 

The following definitions will be useful, as we shall see, in describing the causal PM strategy.

\begin{itemize}
    \item 
    For every $n \in \mbb{N}$, let $\beta(n)$ denote the solution of the following equation
    \begin{align}
    \label{eq:error_exp}
        \beta =\psi^{\ast}(\beta) - \frac{1}{n},
    \end{align}
    where 
    \begin{align}
    \psi(\lambda) := -\log \left\{ (2p)^{\lambda}+(2\p)^{\lambda} \right\} + 1,
    \end{align}
    and  define $\psi^{\ast}(\beta)$ as the Legendre--Fenchel transform of $\psi(\lambda)$:
    \begin{align}
    \label{eq:convex_conj}
        \psi^{\ast}(\beta) := \sup_{\lambda > 0}\left( \psi(\lambda) - \lambda \beta\right).
    \end{align}
    Further, denote by $\lambda^{\ast}(n) \in [0,1]$ the $\lambda$ that achieves the supremum in \eqref{eq:convex_conj} when $\psi^{\ast}(\beta)$ satisfies \eqref{eq:error_exp}.
    
    \item
    For all $i\in \mbb{N}$ and $t \geq n(i-1)+1$, let $k^{(t)}_{i} \in \{0, \ldots, 2^{i}-1\}$ denote the index of the bin containing the median $F^{-1}_{\Theta|Y_1^{t}}\left( {1}/{2}\right)$ in the grid with resolution $2^{-i}$ over the unit interval, i.e, 
    \begin{align}
        k_{i}^{(t)} 2^{-i} \leq F^{-1}_{\Theta|Y^{t}_1} \left( {1}/{2} \middle| y_1^t \right) < \left(k_{i}^{(t)}+1\right) 2^{-i}. 
    \end{align}
\end{itemize}

We are now ready to describe the causal PM strategy in detail.

\subsection{Encoder}
\label{sec:enc}

Fix a parameter $\lambda \in [0,1]$. After $t$ channel uses and the next channel input at time instant $t+1$ is given as follows. Recall that $b(t+1)$ denotes the number of bits available to the encode before transmission at time instant $t+1$. Now, there two cases for the encoding operation:
\begin{itemize}
    \item 
    \underline{No new bit arrives at $t+1$:} In this case, we have $b(t+1) = b(t)$. Hence, the resolution of the grid $2^{-b(t)}$ remains changed. 
    \item
    \underline{A new bit arrives at $t+1$:} In this case, we have $b(t+1) = b(t) + 1$. The encoder divides each of the previous bins into two equal-length bins with equal posterior probabilities in each bin. Hence, the number of bins in the grid increases from $2^{b(t)}$ to $2^{b(t+1)}$. Specifically, for any $k \in \{0, \ldots, 2^i-1\}$ the encoder divides the interval $\left[k2^{-i}, (k+1) 2^{-i}\right)$ into two equal-length bins $\left[(2k) 2^{-(i+1)}, (2k+1) 2^{-(i+1)}\right)$ and $\left[(2k+1)2^{-(i+1)}, (2k+2) 2^{-(i+1)}\right)$ and sets
    \begin{align}
    &\P_{\Theta|Y_1^{t}}\left(\Theta \in \left[(2k) 2^{-(i+1)}, (2k+1) 2^{-(i+1)}\right) \middle| y_1^t \right)
    \\
    &= \P_{\Theta|Y_1^{t}}\left(\Theta \in \left[(2k+1)2^{-(i+1)}, (2k+2) 2^{-(i+1)}\right) \middle| y_1^t \right) 
    \\
    &= \frac{1}{2}\P_{\Theta|Y_1^{t}}\left(\Theta \in \left[k2^{-i}, (k+1) 2^{-i}\right) \middle| y_1^t \right).
    \end{align}
\end{itemize}

Let $d_1^{(t)}$ and $d_2^{(t)}$ denote the values of the probability to the left and to the right of the median in the bin $k_{b(t+1)}^{(t)}$, respectively:
\begin{align}
d_1^{(t)} &:=  {1}/{2} - F_{\Theta|Y^{t}_1} \left(k_{b(t+1)}^{(t)}  2^{-b(t+1)}\right),
\\
d_2^{(t)} &:= F_{\Theta|Y^{t}_1} \left(\left(k_{b(t+1)}^{(t)}+1\right) 2^{-b(t+1)}\right) -  {1}/{2}.
\end{align}
Define further,  
for any $\lambda \in [0,1]$, 
\begin{align}
    \pi_1^{(t+1)}(\lambda) &:= \frac{h(\lambda,d_2^{(t)} )}{h(\lambda,d_1^{(t)} ) + h(\lambda,d_2^{(t)})}, 
    \label{eq:randomization_1}
    \\
  \pi_2^{(t+1)}(\lambda) &:= 1- \pi_1^{(t+1)}(\lambda),
  \label{eq:randomization_2}
\end{align}
where 
\begin{align}
\label{eq:randomization_h}
    h(\lambda, d):= \left( 1 - 2 (\p-p) d \right)^{-\lambda} - \left( 1 +2 (\p-p) d \right)^{-\lambda}.
\end{align}
The next channel input at time $t+1$, conditioned on the past observations $y^{t}_1$, with probability $\pi_1^{(t+1)}(\lambda)$ is given by
\begin{align}
\label{eq:channel_input1}
X_{t+1} =
\left\{
\begin{array}{ll}
0 & \text{if } 0.s_1^{b(t+1)} \leq k_{b(t+1)}^{(t)} 2^{-b(t+1)}, \\
1 & \text{if } 0.s_1^{b(t+1)} > k_{b(t+1)}^{(t)} 2^{-b(t+1)}.
\end{array}
\right.
\end{align}
and with probability $\pi_2^{(t+1)}(\lambda)$ is given by
\begin{align}
\label{eq:channel_input2}
X_{t+1} =
\left\{
\begin{array}{ll}
0 & \text{if } 0.s_1^{b(t+1)} \leq \left(k_{b(t+1)}^{(t)} +1\right)2^{-b(t+1)}, \\
1 & \text{if } 0.s_1^{b(t+1)} > \left(k_{b(t+1)}^{(t)} +1\right) 2^{-b(t+1)}.
\end{array}
\right.
\end{align}

\subsection{Decoder}
\label{sec:dec}
Upon receiving the channel output at time instant $t+1$, the decoder performs a Bayesian update to the posterior of $\Theta$ as follows.

\begin{lemma}
\label{lemma:bayesian_update}
For $t \in \mbb{N}$, consider $i \in \{1,\ldots, b(t+1)\}$. For all $k \leq k_{i}^{(t)}$, since $F_{\Theta|Y_1^{t}} \left(k 2^{-i} \middle| y_1^{t} \right) \leq {1}/{2}$, the Bayesian update after observing the channel output at time $t+1$ is given as follows:
\begin{align}
	\frac{F_{\Theta|Y_1^{t+1}} \left( k 2^{-i} \middle| y_1^{t+1} \right)}{F_{\Theta|Y_1^{t}} \left(k 2^{-i} \middle| y_1^{t} \right)} =
	\begin{cases}
		\frac{p}{\frac{1}{2}+(\p-p)d_1^{(t)}} & \text{if } Y_{t+1} = 1, \\
		\frac{\p}{\frac{1}{2}-(\p-p)d_1^{(t)}} & \text{if } Y_{t+1} = 0 ,
	\end{cases}
\end{align}
with probability $\pi_1^{(t+1)}(\lambda)$ and 
\begin{align}
\frac{F_{\Theta|Y_1^{t+1}} \left( k 2^{-i} \middle| y_1^{t+1} \right)}{F_{\Theta|Y_1^{t}} \left(k 2^{-i} \middle| y_1^{t} \right)} =
\begin{cases}
	\frac{p}{\frac{1}{2}-(\p-p)d_2^{(t)}} & \text{if } Y_{t+1} = 1, \\
	\frac{\p}{\frac{1}{2}+(\p-p)d_2^{(t)}} & \text{if } Y_{t+1} = 0 ,
\end{cases}
\end{align}
with probability $\pi_2^{(t+1)}(\lambda)$. For $k > k_{i}^{(t)} $, since $1-F_{\Theta|Y^{t}}(k2^{-i}) < {1}/{2}$, the Bayesian update for $1-F_{\Theta|Y^{t}}(k2^{-i}) < {1}/{2}$ can be specified similarly. 
\end{lemma}

The proof of Lemma~\ref{lemma:bayesian_update} is given in Appendix~\ref{proof:bayesian_update}.

At any time instant $t$, the decoder generates an estimate $\widehat{\Theta}_t= F^{-1}_{\Theta|Y^{t}}\left( {1}/{2} \middle| y_1^t \right)$ of $\Theta$. The estimates $\hat{s}_1^{b(t)}(t)$ of the first $b(t)$ bits are the binary expansion of the index of the bin containing the median $k_{b(t)}^{(t)}$. Furthermore, when a new bit arrives, similar to the encoder, the decoder divides each bin into two equal-length bins and equally divides the posterior probability.

\begin{remarks}
In the special case where after $t$ channel uses the median coincides with the (left) end point of a bin $k_i^{(t)}2^{-i}$ for some $i \in \{1,\ldots, b(t+1)\}$, at time $t+1$ the encoder transmits $1$ if $s_1^{b(t+1)}$ bits are to the right of the median and---$0$ otherwise. Furthermore, the decoder's update reduces to the update of non-causal PM considered by Horstein in~\cite{Horstein:BSC:IT1963} and Shayevitz and Feder~in~\cite{PosteriorMatching}, where each ${F_{\Theta|Y^{t}}(k 2^{-i})}$, $k \leq k_{i}^{(t)}$ and similarly $1-{F_{\Theta|Y^{t}}(k 2^{-i})}$, $k > k_{i}^{(t)}$, expands by $2\p$ or shrinks by $2p$. 
\end{remarks}

\begin{remarks}
\label{rem:grid_res}
    For any $i\geq 1$, for all $t \geq n(i-1)+1$, 
    the encoder has access to the first $i$ bits and hence the number of bins is at least $2^{i}$. 
    In other words, from time $n(i-1)+1$ to $t$, the causal PM strategy operates on a grid whose resolution is finer than $2^{-i}$ and hence updates $F_{\Theta|Y^{t}}(k2^{-i})$ for all $k \in \{0,\ldots, k_i^{(t)}\}$ and $1- F_{\Theta|Y^{t}}(k2^{-i})$ for all $k \in \{k_i^{(t)}+1, \ldots, 2^{i}-1\}$. 
    This implies that, for all $t \geq n(i-1)+1$, we always encode the first $i$ bits along with the newly available bits. 
\end{remarks}

Although we assume bits arrive one at a time, the strategy and our analysis can be extended to the case where any $m \in \mbb{N}$ bits arrive at a time.

\subsection{Error Exponent Analysis}
\label{sec:main_res}

Now, we provide our main result on the error exponent attained by the causal PM strategy.

\begin{theorem}[Periodic arrival times]
\label{thm:error_exp_fixed_arrival}
Consider the causal PM strategy with parameter $\lambda$ over a $BSC(p)$. The $i$-th bit arrives at the encoder at time $T_i= n(i-1)+1$, i.e., the inter bit-arrival time is constant $n \in \mbb{N}$. Then, for $\lambda = \lambda^{\ast}(n)$, the probability of error in decoding the first $j \in \{1, \ldots, \left\lfloor {t}/{n} \right\rfloor \}$ bits of a message after $t$ channel uses is bounded by
\begin{align}
    \label{eq:error_exp_contant_arr}
    \P\left(\hat{s}_1^j\left(t\right) \neq s_1^j\right) &\leq \kappa \left( 2^{-\beta(n)(t-n(j-1))} \right),
\end{align}
where $\beta(n)$ is the solution of \eqref{eq:error_exp} for $n$, and where $0 \leq \kappa < \infty$. 
\end{theorem}

The proof above theorem is provided in Appendix~\ref{proof:error_exp}. The proof relies on the analyzing the tails of the posterior probability distribution given by $\min\{F_{\Theta|Y_1^{t}} \left( \theta \middle| y_1^t \right), 1-F_{\Theta|Y_1^{t}} \left( \theta \middle| y_1^t \right)\}$ for $\theta \in [0,1)$, which is inspired by the analysis in~\cite{WaeberFrazierHenderson:SICON2013}. However, the analysis of the expected value of decay of the tails is based on the analysis of Burnashev and Zigangirov in~\cite{BurnashevZigangirov1974}.

\section{Applications}
\subsection{Control over Noisy Channels}

\label{ss:NCS}

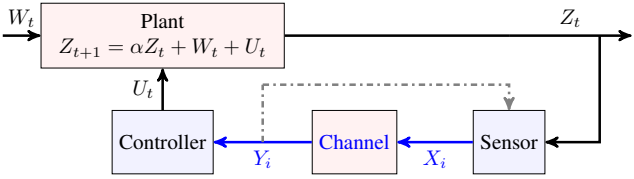
\begin{figure}[t]
    \newcommand{\plant}{Z_{t+1} = \alpha Z_t + W_t + U_t}
    \newcommand{\state}{Z_t}
    \newcommand{\driveNoise}{W_t}
    \newcommand{\contAction}{U_t}
    \newcommand{\channelIn}{X_i}
    \newcommand{\channelOut}{Y_i}
    \scalebox{.8}{\begin{tikzpicture}[auto, arrow/.style={very thick, ->, >=stealth'},node distance=.2\columnwidth,>=latex']
    \node [coord] (input) {};
    \node [plant, right of = input, node distance=.3\columnwidth] (plant) 
    {$\begin{array}{c} 
    \text{Plant} 
    \\ \plant \end{array}$
    };
    \node [coord, right of=plant, node distance = .65\columnwidth] (sum) {};
    \node [coord, right of=sum, node distance = .17\columnwidth] (midoutput) {};
    \node [coord, right of=midoutput, node distance = .06\columnwidth] (output) {};

    \node [block, below of=sum, node distance = .2\columnwidth] (enc) {Sensor};
    \node [block, below of=plant] (dec) {Controller};
    \node [plant, left of=enc, node distance = .29\columnwidth] (channel) {\color{blue} Channel};

    \draw[arrow] (sum) -- node (yArrow) {$\state$} (output);
    \draw[-,very thick] (plant) -- node {} (sum);

    \draw[arrow] (input) -- node {$\driveNoise$} (plant);
    \draw[arrow] (midoutput) |- node[above] {} (enc);
    \draw[arrow] (dec) -- node {$\contAction$} (plant);
    \draw[arrow,color=blue] (enc) -- node {$\channelIn$} (channel);
    \draw[arrow,color=blue] (channel) -- node (channelFB) {$\channelOut$} (dec);
    
    \node [coord, above of=channelFB, node distance = .14\columnwidth] (aboveFB) {};
    \draw[-, very thick, dashdotted, color=gray] (channelFB) |- node {} (aboveFB);
    \node [coord, above of=enc, node distance = .06\columnwidth] (aboveENC) {};
    \draw[arrow, dashdotted, color=gray] (aboveFB) -| node {} (aboveENC);
\end{tikzpicture}
    }
	\caption{A scalar linear plant that is controlled over a noisy channel. The Sensor transmits to the controller over a noisy channel with feedback; $n$ channel uses $\{X_i\}$ per control sample $Z_t$ are assumed.}
    \label{fig:Control_over_BSC_w_FB}
\end{figure}

Consider the problem of stabilizing an unstable scalar plant, 
\begin{align}
\label{eq:control_plant}
Z_{t+1} = \alpha Z_t + W_t + U_t,
\end{align}
where $\alpha > 1$, the initial state is a random variable $Z_0 \in [-\Delta, \Delta]$, the disturbances $\{W_t\}_{t \geq 0}$ are i.i.d.\ with a bounded support $W_t \in [-W,W]$ and $U_t$ is a control signal applied by the controller at time $t$. The controller, that generates $U_t$, is separated from the sensor that measures $Z_t$ by a BSC(p) with feedback, i.e., $n$ channel uses per each control sample $Z_t$ are available. For $\eta \geq 1$, we want to stabilize the $\eta$-th moment, i.e., $\sup_{t}\expe[|Z_t|^{\eta}] <\infty$. To that end, suppose the observer quantizes the plant measurements into 1 bit, which implies a new bit arrives after every $n$ channel uses. This model is depicted in \figref{fig:Control_over_BSC_w_FB}.

\begin{remarks}
For the ease of exposition, 
we consider a 1 bit quantizer but the strategy can be extended to a $k$-bit quantizer for any $1 \leq k \leq n$.
\end{remarks}

To stabilize the plant it suffices to apply a control signal $U_t = -\alpha \hat{Z}_t$, where $\{\hat{Z}_t\}_{t \geq 1}$ satisfies $\sup_{t}\expe[|Z_t-\hat{Z}_t|^{\eta}] < \infty$. The following corollary provides the values $\alpha$ for which the plant can be stabilized. 

\begin{corollary}
\label{coro:max_alpha}
    Consider the plant of equation~\eqref{eq:control_plant} for $\alpha > 1$ observed through a BSC$(p)$ with feedback with a budget of $n$ channel uses. 
    Then, for all $\eta \geq 1$, the plant is $\eta$-stabilizable, i.e., 
    $\sup_{t} \expe\left[\left|Z_t - \hat{Z}_t \right|^{\eta}\right] < \infty$, 
    for
    \begin{align}
    \label{eq:max_alpha}
        \log \alpha \leq \min\left\{\frac{1}{n}, \frac{\beta(n)}{\eta}\right\},
    \end{align} 
    where $\beta(n)$ is the solution of equation~\eqref{eq:error_exp}.
\end{corollary}

\begin{IEEEproof}
We use the causal PM strategy to transmit the quantized plant measurements over a BSC$(p)$ with feedback. We apply our causal PM strategy for the case where inter-arrival time of the bits is a constant $n$, hence we set $\lambda = \lambda^{\ast}(n)$. For each step of the plant evolution we convey one bit over $n$ channel uses. Theorem~\ref{thm:error_exp_fixed_arrival} provides the following guarantees on the estimates generated by the causal PM~strategy
\begin{align}
&\P\left(\hat{s}_1^j\left(nt\right) \neq s_1^j\right)
\nonumber
\leq 
\kappa \left( 2^{-\beta(n)n(t-j)} \right),
\end{align}
for all $j \in \{1,\ldots, t\}$, where $\beta(n)$ is the solution of equation~\eqref{eq:error_exp}. Hence, using \cite[Theorem~4.1]{SahaiMitterPartI} we have that the plant is $\eta$-stabilizable if equation~\eqref{eq:max_alpha} holds.
\end{IEEEproof}

\begin{remarks}
The constraint $\log \alpha < 1/n \leq 1$ is due to a 1-bit quantization requirement that we implicitly impose by assuming that a single bit arrives at a time. This requirement can be lifted by allowing higher quantization rates, along with the appropriate adaptation of the proposed scheme, at the price of reducing the error exponent $\beta$. In other words, two conflicting effects can be seen in the problem of stabilizing an unstable plant over a noisy channel: (i) Source quantization: we wish to maximize the quantization resolution to allow for finer source approximation, however this results in higher channel-coding rate since more bits have to be sent over a given channel budget $n$. (ii) Channel coding: we wish to minimize the channel-coding rate to minimize the error due to decoding, i.e., to maximize the error exponent. These two effects are manifested by the two minimands in equation~\eqref{eq:max_alpha}.
\end{remarks}

\begin{remarks}
As a consequence of Theorem~\ref{thm:error_exp_fixed_arrival}, for a given $\eta \geq 1$ and $p \in (0, 1/2)$, we obtain a lower bound $R(p)$ on the maximum rate (i.e., minimum channel budget $\lceil 1/{R(p)} \rceil$) at which the communication channel BSC$(p)$ can be operated such that the plant~\eqref{eq:control_plant} is $\eta$-stabilizable for some $\alpha > 1$. Using equation~\eqref{eq:error_exp}, note that we have
\begin{align}
    &\min\left\{ \frac{1}{n}, \frac{\beta(n)}{\eta} \right\} 
    \geq \max_{\beta > 0} \min \left\{ \frac{\beta}{\eta}, \frac{1}{\eta} \left( \psi^{\ast}(\beta)-\frac{1}{n} \right), \frac{1}{n} \right\}
\nonumber
\end{align}
Hence, using \colref{coro:max_alpha}, this implies that $R(p)$ is the largest $R > 0$ that satisfies the following equation:
\begin{align}
    \psi^{\ast}(\eta R)= (\eta +1)R.
\end{align}
In other words, we obtain that $2^{R(p)}$ is a lower bound on the largest $\alpha$ for which the plant~\eqref{eq:control_plant} can be $\eta$-stabilized over a BSC$(p)$ for any channel budget $n > 1$.
\end{remarks}


\subsubsection{Simulations for Control over Noisy Channels}
\label{sec:numeric}

\begin{figure}[t]
\centering
\vspace{-\baselineskip}
    \includegraphics[width=0.9\columnwidth]{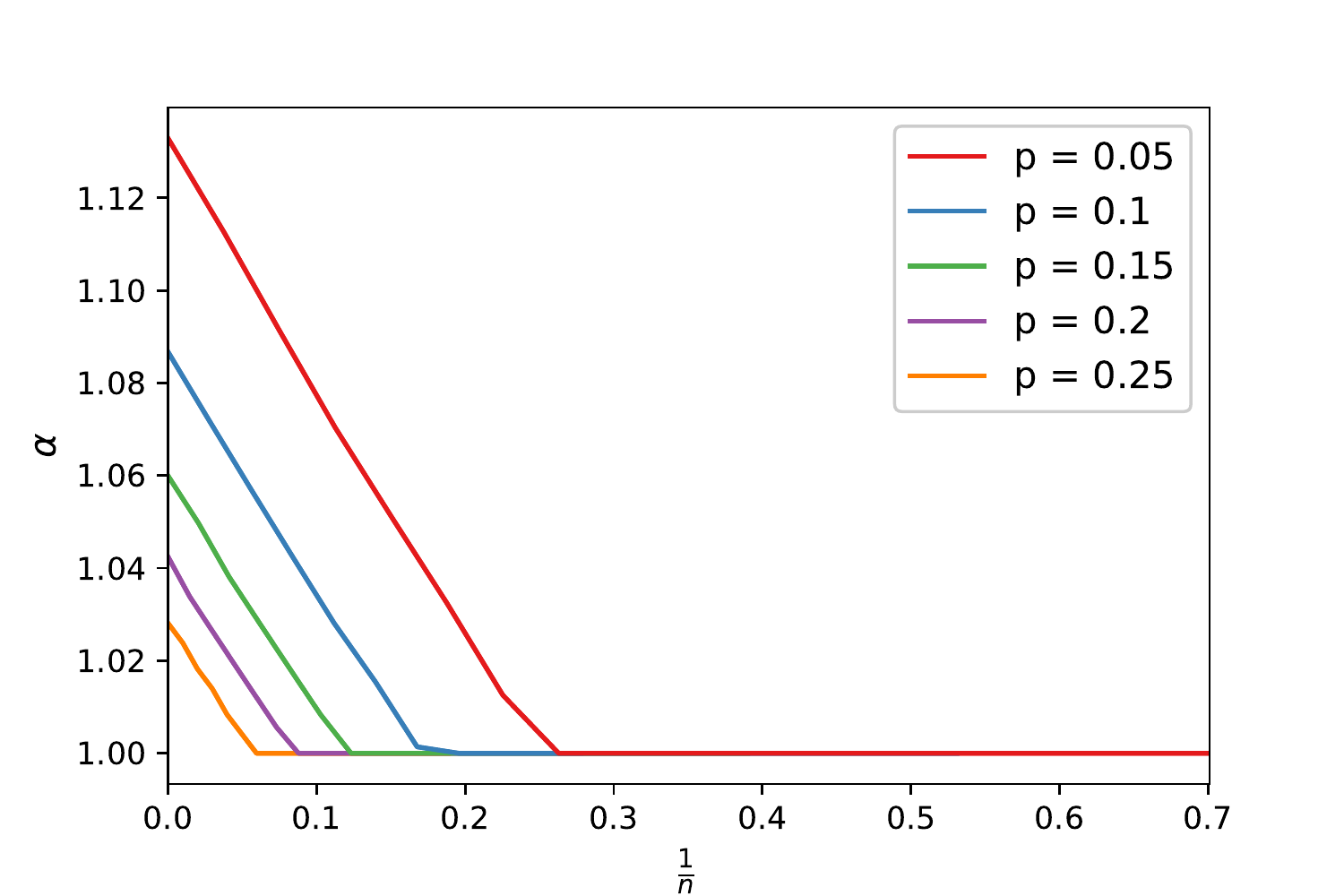}
    \caption{The maximum eigenvalue $\alpha$ of a plant that is stabilizable using the causal PM strategy over a BSC$(p)$ as a function of inverse of channel budget $1/n$ for various values of $p$.}
\label{fig:alphas_vs_channel_budget}
\end{figure}

For various values of channel budget $n \in \mbb{N}$ we numerically compute the bound provided by Corollary~\ref{coro:max_alpha} on largest eigenvalue $\alpha$ for which a plant is stabilizable. \figref{fig:alphas_vs_channel_budget} shows the largest eigenvalue $\alpha$ as a function of inverse of the channel budget, i.e. rate, for different values of crossover probability of a BSC. This illustrates that the causal PM-based scheme can stabilize the plant for $\alpha$ values that are strictly greater than one for the considered crossover probabilities by choosing channel budget appropriately.

\begin{figure}[t]
\centering
    \includegraphics[width=0.8\columnwidth, trim = {8mm 0 8mm 4mm}, clip]{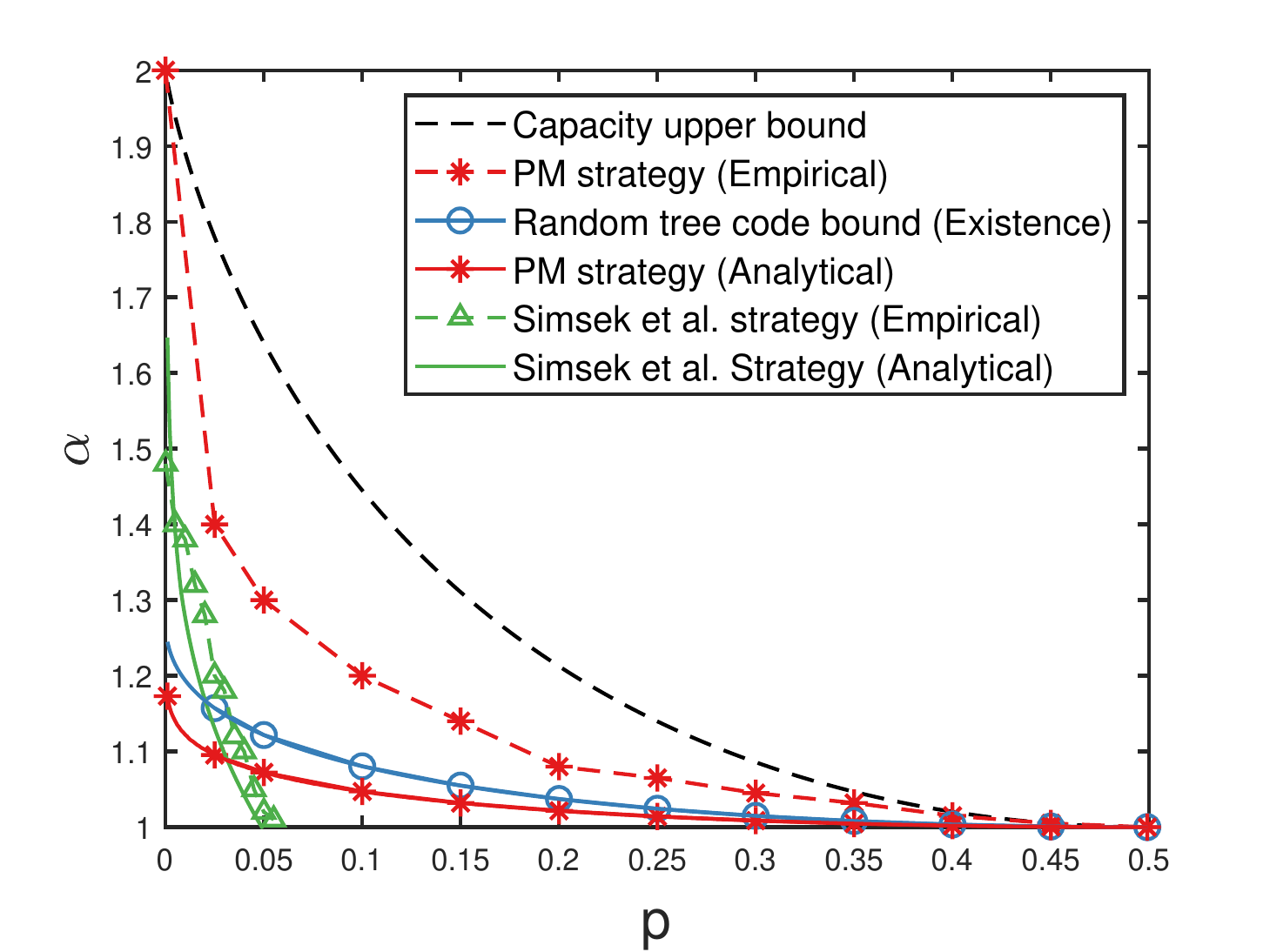}
    \caption{The maximum eigenvalue $\alpha$ of a plant that is stabilizable over a BSC$(p)$ as a function of $p$ using:
    the causal PM strategy (analytically and empirically), the Simsek et al. strategy~\cite{SimsekJainVaraiya:AC2004} (analytically and empirically), the Sahai--Mitter tree-code lower bound \cite{SahaiMitterPartI}, and the capacity upper bound.} 
\label{fig:alphas}
\end{figure}

We compare the performance of the proposed causal PM strategy with previously proposed upper and lower bounds for the maximal value of $\alpha$  for which the plant~\eqref{eq:control_plant} can be stabilized.
\figref{fig:alphas} compares the stabilizability of a system as a function of the crossover probability of a BSC. 
The empirical as well as the theoretical performance of both the causal PM-based strategy and a strategy proposed by Simsek et al. \cite{SimsekJainVaraiya:AC2004} (albeit for the interference-free case: $W_t \equiv 0$), as well as the Sahai--Mitter lower bound without feedback (anytime-reliable tree codes) of~\cite{SahaiMitterPartI} and the capacity upper bound are illustrated. 
From \figref{fig:alphas} we see that the bound $2^{R(p)}$ on $\alpha$, provided by our analysis of the causal PM-based scheme, is rather conservative in comparison to its empirical performance. The latter clearly outperforms the Simsek et al.~strategy~\cite{SimsekJainVaraiya:AC2004} (for which the analysis is rather tight) and exceeds the Sahai--Mitter lower bound. This demonstrates that the causal PM-based strategy 
provides better performance both in terms of stability and complexity.
We further note that the causal PM-based scheme can stabilize the plant for $\alpha$ values that are strictly greater than one for all crossover probabilities $p \in [0, 1/2)$, 
even under the provided conservative analysis.
This is in stark contrast to the strategy of Simsek et al., which can stabilize unstable plants only below a certain threshold crossover probability.

\subsection{Streaming Bits with Random Inter Bit-Arrival Times}
In this section, we assume that the bits of the message point $\Theta$ are revealed to the transmitter causally at arbitrary (possibly random) times, as follows. Let $\{N_i\}_{i \geq 1}$ be an i.i.d.\ random process where each $N_i$ has a pmf $p_N$ and $N_i \in  [n_{\min}, n_{\max}]$, where $n_{\min} < n_{\max}$, and $n_{\min}, n_{\max} \in \mbb{N}$. Furthermore, the $i$-th bit arrives at time $T_i := \sum_{j=1}^{i-1}N_j+1$ for all $i \geq 2$ with $T_1 = 1$. For all time instants $t \in \mbb{N}$, recall that $b(t):= \max\{i \in \mbb{N}: T_{i} \leq t\}$. Assuming that the first bit is available at the beginning (i.e., $T_1 = 1$), at time $t$ at most first $\lceil\frac{t}{n_{\min}}\rceil$ bits have arrived with a non-zero probability and similarly the first $\lceil \frac{t}{n_{\max}}\rceil$ bits have arrived with probability 1, which implies
\begin{align}
  \P \left( 
  \left\lceil \frac{t}{n_{\max}}\right\rceil 
  \leq b(t) \leq 
  \left\lceil\frac{t}{n_{\min}}\right\rceil 
  \right) = 1 .
\end{align}

\begin{figure}[!htb]
\centering
    \includegraphics[width=.94\columnwidth]{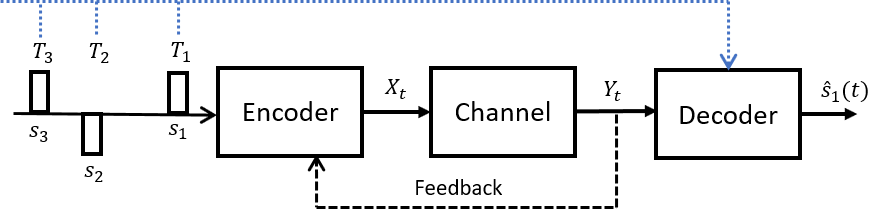}
    \caption{Figure shows the transmission of a stream of bits which arrive at the encoder at random times $T_i$. The arrival times are available at both at the encoder and decoder.}
\label{fig:channel_block_diag}
\end{figure}

Note that deterministic and periodic arrival times, in which a new information bit is revealed every fixed $n \in \mbb{N}$ time step is a special case of this framework. We now define the feedback communication scheme of an information bit sequence that is made available causally to the encoder with random inter bit-arrival times, depicted in Figure.~\ref{fig:channel_block_diag}. We assume that the times at which the bits are revealed to the transmitter are known at the receiver. 

Causal posterior matching strategy described in Section~\ref{sec:PM} can be applied to this by setting the parameter $\lambda \in \left\{\lambda^{\ast}(n_{\min}), \lambda^{\ast}(n_{\max})\right\}$. Next, we provide error exponents achieved by causal PM strategy when information bit sequence has random inter bit-arrival times.

\begin{theorem}
\label{thm:error_exp_random_arrival}
    Consider the causal PM strategy with parameter $\lambda$ over a $BSC(p)$. The $i$-th bit arrives at the encoder at a random time $T_i$, whose pmf is $p_N^{\bigotimes i}$,
    where the inter-arrival times lie in the set $ [n_{\min}, \ldots, n_{\max}]$ and  let
    \begin{align*}
        b(t) = \max\{i \in \mbb{N}: T_i \leq t\}.
    \end{align*}
    Then, 
    \begin{enumerate}[(i)]
    \item 
    \label{itm:regime:high}
        For $\lambda = \lambda^{\ast}(n_{\min})$, the probability of error in decoding the first $i \in \{1, \ldots, \left\lceil {t}/{n_{\min}} \right\rceil \}$ message bits after $t$ channel uses is bounded by 
        \begin{align}
        \label{eq:error_exp_min}
            &\P\left(\left.\hat{s}_1^i\left(t\right) \neq s_1^i\right|b(t) > i\right)
            \nonumber
            \\
            &\leq \kappa \expe\left[\left. 2^{-\beta(n_{\min})(t-T_{i})}\right| b(t) > i\right], 
        \end{align}
        where $\beta(n_{\min})$ is the solution of equation~\eqref{eq:error_exp} for $n = n_{\min}$ and where $\kappa$ is a finite positive constant.
    \item 
    \label{itm:regime:low}
        For $\lambda = \lambda^{\ast}(n_{\max})$, the probability of error in decoding the first $i \in \{1, \ldots, \left\lfloor {t}/{n_{\max}} \right\rfloor \}$ message bits after $t$ channel uses is bounded by
        \begin{align}
        \label{eq:error_exp_max}
            \P\left(\hat{s}_1^i \left(t\right) \neq s_1^i\right) &\leq \kappa  2^{-\beta(n_{\max})(t-n_{\max}(i-1))} ,
        \end{align}
        where $\beta(n_{\max})$ is the solution of equation~\eqref{eq:error_exp} for $n = n_{\max}$ and where $\kappa$ is a finite positive constant.
    \end{enumerate}
\end{theorem}

The proof above theorem is provided in Appendix~\ref{proof:error_exp_random}.

Theorem~\ref{thm:error_exp_random_arrival} shows that the causal PM strategy can operate in two regimes based on how the randomization probabilities $\pi_1^{(t)}$ and $\pi_2^{(t)}$ are chosen given the past observations and the number of bits available at the encoder, i.e., by setting the parameter $\lambda$ appropriately. In the setting~(\ref{itm:regime:high}), causal PM can be thought of as operating in a ``\textit{high-rate regime}", 
since it decodes all the arrived information bits, 
but with a lower error exponent of equation~\eqref{eq:error_exp_min}, corresponding to $\beta(n_{\min})$. 
In contrast, in the setting~(\ref{itm:regime:low}), 
causal PM can be thought of as operating in a ``\textit{low-rate regime}", as it decodes only the first $\left\lfloor {t}/{n_{\max}} \right\rfloor$ bits, but with a higher error exponent of equation~\eqref{eq:error_exp_max}, corresponding to $\beta(n_{\min})$. 

\section{Conclusions and Future Work}
\label{sec:summary}

We considered the problem of transmitting an infinite stream of bits over a BSC where the bits are revealed to~the transmitter causally and the inter bit-arrival time may be random. We proposed a causal PM strategy and provided guarantees for the error exponent of the decoded bits using this strategy. The causal PM is parameterized by $\lambda(n)$ which decides the randomization of the encoding functions. Hence, it implicitly decides the number of bits decoded and their error exponent. We derived explicit results for two extremes of $\lambda(n)$. 
An interesting area of future work would be to extend our analysis to any $\lambda$ between these two extremes. Another important future direction is to extend our analysis to the case where the bit arrival times are unknown at the receiver. 

Furthermore, we applied our strategy to the problem of stabilizing a control plant over a BSC. We provided analytical guarantees on the maximal plant eigenvalue for which the plant can be stabilized using causal posterior matching. Closing the gap between our analysis and the empirical performance is an important area of future.

\appendix

\subsection{Preliminaries}

\subsubsection{Proof of Lemma~\ref{lemma:bayesian_update}}
\label{proof:bayesian_update}
For $k \in \{0, \ldots, 2^{i}-1\}$, note that $F_{\Theta|Y_1^{t}}(k2^{-i}|y_1^{t}) = \P_{\Theta|Y_1^{t}}(\Theta \in [0, k 2^{-i}) \mid y_1^t)$. For any $y \in \{0,1\}$ after observing $Y_{t+1} = y$, the decoder updates the posterior distribution of $\Theta$ using the Bayes rule as follows 
\begin{align}
    &F_{\Theta|Y_1^{t+1}}(k2^{-i}|y_1^{t}, Y_{t+1} = y)
    \\
    &= 
    \frac{F_{\Theta|Y_1^{t}}(k2^{-i}|y_1^{t})\P(Y_{t+1} = y\mid \Theta \in [0, k2^{-i}], y_1^t) }{\P(Y_{t+1} = y \mid y^{t}_1)}.
\end{align}

Let $Y_{t+1} = 1$ and the case where $Y_{t+1} = 0$ can be obtained similarly. First consider the case where median is approximated as the left end point of the $k_{b(t+1)}^{(t)}$th i.e., $k_{b(t+1)}^{(t)}2^{-b(t+1)}$. Then recall that the encoding function is given by equation~\eqref{eq:channel_input1}. For all $i \in \{1,\ldots, b(t+1)\}$ since
\begin{align}
    k_i^{(t)}2^{-i} \leq k_{b(t+1)}^{(t)}2^{-b(t+1)}
\end{align} 
we have $\P(Y_{t+1} = 1\mid \Theta \in [0, k2^{-i}], y_1^t)  = p$ for all $k \in \{0,\ldots, k_i^{(t)}\}$. Furthermore, we have
\begin{align}
    &\P(Y_{t+1} = 1 \mid y^{t}_1)
    \\
    &= 
    \P(Y_{t+1} = 1 \mid \Theta \in [0, k_{b(t+1)}^{(t)}2^{-b(t+1)}), y_1^t) \times
    \\
    &\hspace{0.2cm}\times F_{\Theta|Y_1^t}(  k_{b(t+1)}^{(t)}2^{-b(t+1)} \mid y_1^{t})
    \\
    & + \P(Y_{t+1} = 1 \mid \Theta \in [ k_{b(t+1)}^{(t)}2^{-b(t+1)}, 1), y_1^t) \times
    \\
    &\hspace{0.2cm}\times (1- F_{\Theta|Y_1^t}(  k_{b(t+1)}^{(t)}2^{-b(t+1)} \mid y_1^{t}))
    \\
    &= p \left(\frac{1}{2}-d_1^{(t)}\right) + \p\left(\frac{1}{2}+d_1^{(t)}\right)
    \\
    & = \frac{1}{2} + (\p - p) d_1^{(t)}.
\end{align}
Hence, for all $k \in \{0, \ldots, k_i^{(t)}\}$ we obtain 
\begin{align}
    F_{\Theta|Y_1^{t+1}}(k2^{-i}|y_1^{t+1})
    = 
    \frac{F_{\Theta|Y_1^{t}}(k2^{-i}|y_1^{t}) p}{\frac{1}{2} + (\p - p) d_1^{(t)}}. 
\end{align}
Similarly we can obtain the Bayes update for $1-F_{\Theta|Y_1^{t+1}}(k2^{-i}|y_1^{t+1})$ when $k \in \{k_i^{(t)}+1, \ldots, 2^{i}-1\}$. Now consider the case where median is approximated as the right end point of the $k_{b(t+1)}^{(t)}$th i.e., $(k_{b(t+1)}^{(t)}+1) 2^{-b(t+1)}$. Then recall that the encoding function is given by equation~\eqref{eq:channel_input2}. Again noting that 
\begin{align}
        &k_i^{(t)}2^{-i} 
        \leq k_{b(t+1)}^{(t)}2^{-b(t+1)}  
        \\
        &(k_{b(t+1)}^{(t)}+1) 2^{-b(t+1)}  
        \leq (k_i^{(t)} +1)2^{-i}
\end{align} 
we have $\P(Y_{t+1} = 1\mid \Theta \in [0, k2^{-i}], y_1^t)  = p$ for all $k \in \{0,\ldots, k_i^{(t)}\}$ and $\P(Y_{t+1} = 1 \mid y^{t}_1) = \frac{1}{2} - (\p - p) d_2^{(t)}$. The Bayesian update for the rest of the all cases can be obtained similarly. Note that the proof relies on the fact that $i \leq b(t+1)$, hence for bits which have not arrived by $t+1$ the update of the posterior probabilities may not be updated according to the assertion of the lemma.

\subsection{Proof of Theorem~\ref{thm:error_exp_fixed_arrival}}

\label{proof:error_exp}

Consider time instants $t \geq n(i-1)+1$ and hence, the encoder has access to first $i$ bits. Recall that $k^{(t)}_{i}$ denotes the index of the bin containing the median $F^{-1}_{\Theta|Y^{t}}\left( \frac{1}{2}\right)$ in grid with resolution $2^{-i}$ i.e., 
\begin{align}
k^{(t)}_{i} 2^{-i} \leq F^{-1}_{\Theta|Y^{t}}\left( \frac{1}{2}\right) < (k^{(t)}_{i}+1) 2^{-i}. 
\end{align}
Consider the probability of having an error in the first $i$ bits and we get
\begin{align}
&\P\left(\hat{s}_1^i \left(t\right) \neq s_1^i \right)
\\
&=
\P\left(  \Theta \not \in \left(k^{(t)}_{i} 2^{-i}, (k^{(t)}_{i}+1) 2^{-i} \right] \right)
\\
& = 
\expe\left[F_{\Theta|Y^{t}} \left(k^{(t)}_{i}2^{-i}\right) 
+ 1 - F_{\Theta|Y^{t}}\left( (k^{(t)}_{i}+1) 2^{-i}\right) \right].
\end{align}
For any $\tau \geq 1$, define
\begin{align}
\xi_{\Theta|Y^{\tau}}(x):= \min\left\{F_{\Theta|Y^{\tau}}(x), 1- F_{\Theta|Y^{\tau}}(x)\right\}    
\end{align}
where $x = k2^{-i}$ and $k\in \{0, \ldots, 2^{i}-1\}$. Since $F_{\Theta|Y^{t}} \left( k_{i}^{(t)} 2^{-i}\right) \leq \frac{1}{2}$ and $1 - F_{\Theta|Y^{t}}\left( (k_{i}^{(t)}+1) 2^{-i}\right) \leq \frac{1}{2}$, note that $\xi_{\Theta|Y^{t}}\left(k_{i}^{(t)} 2^{-i}\right) = F_{\Theta|Y^{t}} \left( k_{i}^{(t)}2^{-i}\right)$ and that $\xi_{\Theta|Y^{t}} \left(  (k_{i}^{(t)} +1)2^{-i}\right) = 1-F_{\Theta|Y^{t}}\left( (k_{i}^{(t)} +1)2^{-i}\right)$. Hence, for every $\beta > 0$  we can write
\begin{align}
&\P\left(\hat{s}_1^i \left(t\right) \neq s_1^i\right)
\\
& = 
\expe\left[ \xi_{\Theta|Y^{t}} \left(k^{(t)}_{i}2^{-i}\right) + \xi_{\Theta|Y^{t}}\left( (k^{(t)}_{i} +1)2^{-i}\right) \right]
\\
&\leq
2^{-\beta (t- n(i-1))} + \P\left(  \xi_{\Theta|Y^{t}}\left( (k^{(t)}_{i}+1) 2^{-i}\right)  \right.+
\\
&
+\left. \xi_{\Theta|Y^{t}} \left(k^{(t)}_{i} 2^{-i}\right) \geq 2^{-\beta (t- n(i-1))}\right).
\label{eq:prob_error_random_arr}
\end{align}

The second term in the above equation can be bounded  using Markov's inequality for any $\lambda > 0$ as follows
\begin{align}
&\P\left(  \xi_{\Theta|Y^{t}} \left((k^{(t)}_{i}+1)2^{-i}\right) + \xi_{\Theta|Y^{t}}\left( k^{(t)}_{i} 2^{-i}\right)\right.
\\
&\hspace{1cm}\left.\geq 2^{-\beta (t- n(i-1))} \right)
\\
&\leq 2^{\lambda\beta (t-n(i-1))}\times
\\ 
& \hspace{0.5cm}\expe\left[\left(\xi_{\Theta|Y^{t}} \left(k^{(t)}_{i}2^{-i}\right) + \xi_{\Theta|Y^{t}}\left( (k^{(t)}_{i}+1) 2^{-i}\right)\right)^{\lambda}\right].
\label{eq:markov_ineq}
\end{align}
Furthermore, note that
\begin{align}
&\expe\left[\left(\xi_{\Theta|Y^{t}} \left(k^{(t)}_{i}2^{-i}\right)+\xi_{\Theta|Y^{t}}\left( (k^{(t)}_{i}+1) 2^{-i}\right)\right)^{\lambda}\right]
\\
& 
\leq 
2^{\lambda}\expe\left[ 
\max_{k}\xi^{\lambda}_{\Theta|Y^{t}}\left( k 2^{-i}\right)
\right]
\nonumber
\\
& \leq 
2^{\lambda}\sum_{k=0}^{ 2^{i}-1}\expe\left[
\xi^{\lambda}_{\Theta|Y^{t}}\left( k 2^{-i}\right)
\right].
\label{eq:sum_of_tails_random_arr}
\end{align}

Fix some $k \in \{0,\ldots, 2^i-1\}$. Then, for any  $n(i-1)+1\leq \tau \leq t$, consider
\begin{align}
&\expe\left[ \xi^{\lambda}_{\Theta|Y^{\tau}}(k2^{-i}) \right]
\\
&= 
\expe\left[2^{{\lambda}\log \xi_{\Theta|Y^{\tau}}(k2^{-i}) } \right]
\\
& = 
\expe\left[ 2^{\lambda \log\xi_{\Theta|Y^{\tau-1}}(k2^{-i})} \left.\expe\left[ 2^{\lambda\log \frac{\xi_{\Theta|Y^{\tau}}(k2^{-i})}{\xi_{\Theta|Y^{\tau-1}}(k2^{-i})} } \right| Y^{\tau-1}  \right]\right]
\\
& \overset{(a)}\leq 
2^{-\psi(\lambda)} \expe\left[ 2^{\lambda\log\xi_{\Theta|Y^{\tau-1}}(k2^{-i})}\right],
\end{align}
where $\psi(\lambda) = -\log \left( \frac{  (2p)^{\lambda}+(2\p)^{\lambda}}{2} \right)$ and $(a)$ obtained by applying Lemma~\ref{lemma:mgf_bound}.

For all $\tau \geq n(i-1)+1$ the causal PM strategy operates on a grid whose resolution is finer than $2^{-i}$. As discussed in Remark~\ref{rem:grid_res} this implies that $\xi_{\Theta|Y^{\tau}}(k2^{-i})$ is updated for all $n(i-1)+1 \leq \tau \leq t$. Therefore, applying Lemma~\ref{lemma:mgf_bound} repeatedly for all time instants $n(i-1)+1 \leq \tau \leq t$, we obtain the following for every $k \in \{0, \ldots, 2^{i}-1\}$
\begin{align}
&\expe\left[ \xi^{\lambda} _{\Theta|Y^{t}}(k2^{-i}) \right]
\\
&\leq 
2^{-\psi(\lambda) (t-n(i-1))} \expe\left[  \xi^{\lambda}_{\Theta|Y^{n(i-1)}}(k2^{-i}) \right].
\label{eq:avg_single_tail_random_arr}
\end{align}
Substituting equation~\eqref{eq:avg_single_tail_random_arr}  in equation~\eqref{eq:sum_of_tails_random_arr} we have
\begin{align}
&\expe\left[\left(\xi_{\Theta|Y^{t}} \left(k^{(t)}_{i}2^{-i}\right)+\xi_{\Theta|Y^{t}}\left( (k^{(t)}_{i}+1) 2^{-i}\right)\right)^{\lambda}\right]
\\
&\leq 2. 2^{-\psi(\lambda) (t-n(i-1))} \sum_{k=0}^{2^i-1} \expe\left[\xi^{\lambda}_{\Theta|Y^{n(i-1)}}(k2^{-i}) \right].
\end{align}
Now, applying Lemma~\ref{lemm:posterior_shape_random_arr} for $\lambda$ such that $\psi(\lambda)-\frac{1}{n} > 0$ we have
\begin{align}
&\expe\left[\left(\xi_{\Theta|Y^{t}} \left(k^{(t)}_{i}2^{-i}\right) +\xi_{\Theta|Y^{t}}\left( (k^{(t)}_{i} +1)2^{-i}\right)\right)^{\lambda}\right]
\\
& \leq  2.2^{-\psi(\lambda) (t-n(i-1))}
\frac{1}{1-2^{-\left(\psi(\lambda)-\frac{1}{n}\right)n}}.
\end{align}
Substituting the above inequality in equation~\eqref{eq:markov_ineq} followed by substituting this in equation~\eqref{eq:prob_error_random_arr}, the probability of error in decoding first $i$ bits for all $\lambda > 0$ such that $\psi(\lambda)-\frac{1}{n} > 0$ is given by
\begin{align}
&\P\left(\hat{s}_1^i\left(t\right) \neq s_1^i \right)
\\
&\leq 
2^{-\beta (t-n(i-1))}+  
\frac{2.2^{-\left(\psi(\lambda)-\lambda \beta \right)(t-n(i-1))}}{1-2^{-\left(\psi(\lambda)-\frac{1}{n}\right)n}}.
\end{align}

For any $\beta > 0$, if there exists a $\lambda > 0$ such that $\psi(\lambda)-\lambda \beta - \frac{1}{n} > 0$, then we have $\psi(\lambda) - \frac{1}{n} > 0$ and $\psi(\lambda)-\lambda \beta > 0$. Also, if $\sup_{\lambda > 0}(\psi(\lambda)-\lambda \beta)-\frac{1}{n} > 0$ then the supremum achieving $\lambda$ also satisfies $\psi(\lambda)-\frac{1}{n} > 0$. Therefore, we fix $\lambda = \lambda^{\ast}(n)$. Hence, we obtain the following
\begin{align}
&\P\left(\hat{s}_1^i\left(t\right) \neq s_1^i\right)
\\
&\leq
2^{-\beta(t-n(i-1))} +
\kappa^{\prime}2^{-\left(\psi^{\ast}(\beta) -\frac{1}{n}\right)(t-n(i-1))}
\\
& \leq
\kappa 2^{-\max_{\beta > 0}\min\left\{\beta, \psi^{\ast}(\beta)-\frac{1}{n}\right\} (t-n(i-1))},
\\
& \leq 
\kappa 2^{-\beta(n) (t-n(i-1))}.
\end{align}  
for some positive constants $\kappa^{\prime},\kappa < \infty$ independent of $t$.

\subsection{Proof of Theorem~\ref{thm:error_exp_random_arrival}}

\label{proof:error_exp_random}
Consider the sample path where the inter-bit arrival realizations $\{N_{j}\}_{j \geq 1}$ are $ \{n_j\}_{j \geq 1}$. Let $t_{i}: = \sum_{j=1}^{i-1}n_j +1$. Fix $i \in \{1,\ldots, \frac{t}{n_{\min}}\}$ and consider the event where $T_{i} = t_i < t$. This is the case where first $i$ bits arrive by time instant $t$ with non-zero probability. The probability of error in decoding first $i$ bits conditioned on the event $\{N_1^{i-1} = n_1^{i-1}\}\cap \{T_i < t\}$ can be upper bounded as follows by setting $\lambda = \lambda^{\ast}(n_{\min})$ applying Theorem~\ref{thm:error_exp_fixed_arrival}:
\begin{align}
&\P\left(\left.\hat{s}_1^i\left(t\right) \neq s_1^i\right| N_1^{i-1} = n_1^{i-1},  T_i < t\right)
\\
& \leq 
\kappa 2^{-\beta(n_{\min}) (t-t_{i})}.
\end{align} 
Now, note that the event $\{N_1^{i-1} = n_1^{i-1}\}\cap \{T_i < t\}$ is equivalent to the event $\{N_1^{i-1} = n_1^{i-1}\}\cap \{b(t) > i\}$. Taking the expectation with respect to $p_N^{\bigotimes i}$ we have
\begin{align}
    \P\left(\left.\hat{s}_1^i\left(t\right) \neq s_1^i\right|b(t) > i\right)
    \leq \kappa \expe\left[\left. 2^{-\beta(n_{\min})(t-T_{i})}\right| b(t) > i\right], 
    \end{align}
for all $i \in \left\{1,\ldots, \left\lceil \frac{t}{n_{\min}}\right\rceil\right\}$.

Set $\lambda = \lambda^{\ast}(n_{\max})$. Now, following the same steps as in the proof of Theorem~\ref{thm:error_exp_fixed_arrival} where instead of Lemma~\ref{lemm:posterior_shape_random_arr} we use Lemma~\ref{lemm:posterior_shape}, for $i \in \{1, \ldots, \lceil \frac{t}{n_{\max}}\rceil\}$, we obtain the following
\begin{align}
&\P\left(\hat{s}_1^i\left(t\right) \neq s_1^i\right)
\\
& \leq
\kappa 2^{-\max_{\beta > 0}\min\{\beta, \psi^{\ast}(\beta)-\frac{1}{n_{\max}}\} (t-n_{\max}(i-1))}
\\
&= \kappa 2^{-\beta(n_{\max})(t-n_{\max}(i-1))},
\end{align}  
some positive constant $\kappa < \infty$ independent of $t$.

\subsection{Technical Background}
In this appendix, we provide some preliminary lemmata which are technical and only helpful in proving the main results of the paper. Furthermore, we provide a general version of the lemmata for bits with random but bounded inter bit-arrival times. 

\begin{lemma}
\label{lemma:mgf_bound}
For $t \geq 1$ consider $i \in \left[\left\lceil\frac{t}{n_{\min}} \right\rceil \right]$. Consider a sample path where $N_1^{i-1} = n_1^{i-1}$ and $b(t) > i$. Then for all $\tau$ such $t_{i} - 1= \sum_{j=1}^{i-1}n_j \leq \tau \leq t$, for any point $x = k2^{-i}$, where $k \in \{0,\ldots, 2^{i}-1\}$, and $0 < \lambda \leq 1$ the following holds true
\begin{align}
&\expe\left[ \left.\left(\frac{\xi_{\Theta|Y^{\tau+1}}(x)}{\xi_{\Theta|Y^{\tau}}(x) }\right)^{\lambda} \right| Y^{\tau}, N_1^{i-1}= n_1^{i-1}, T_i < t \right] 
\\
&\leq \frac{ (2p)^{\lambda} + (2\p)^{\lambda}}{2},
\end{align}
when randomization probabilities are chosen as follows
\begin{align}
&\pi_1^{(\tau+1)}(\lambda)
= 
\frac{h(\lambda,d_2^{(\tau)} )}{h(\lambda,d_1^{(\tau)} ) + h(\lambda,d_2^{(\tau)} )},
\\
&\pi_2^{(\tau+1)}(\lambda) = 1- \pi_1^{(\tau+1)}(\lambda),
\end{align}
where $d_1^{(\tau)}$ and $d_2^{(\tau)}$ are as defined in equation~\eqref{eq:randomization_1} respectively and \eqref{eq:randomization_2} and $h(\lambda, d)$ is defined in equation~\eqref{eq:randomization_h}.
\end{lemma}

\begin{IEEEproof}
The proof of this lemma is based on the analysis of moment generated function provided by Burnashev and Zigangirov in~\cite{BurnashevZigangirov1974}.

\noindent
\underline{Case 1:} Suppose the median does not cut any bin i.e., median coincides with the end point of some bin. 

\underline{Case 1a:}
Using Lemma~\ref{lemma:bayesian_update}, the update of $\xi_{\Theta|Y^{\tau}}(x)$ when $F_{\Theta|Y^{\tau}}(x) \leq \frac{1}{2}$ is given as follows 
\begin{align}
\frac{\xi_{\Theta|Y^{\tau+1}}(x)}{\xi_{\Theta|Y^{\tau}}(x)} \leq
\left\{
\begin{array}{lll}
 2p & \text{if } Y_{\tau+1} = 1 &\text{ w.p. }\frac{1}{2},\\
 2\p & \text{if } Y_{\tau+1} = 0 &\text{ w.p. }\frac{1}{2}.
\end{array}
\right.
\end{align}

\underline{Case 1b:} Using Lemma~\ref{lemma:bayesian_update}, the update of $\xi_{\Theta|Y^{\tau}}(x)$ when $1-F_{\Theta|Y^{\tau}}(x) < \frac{1}{2}$ is given as follows 
\begin{align}
\frac{\xi_{\Theta|Y^{\tau+1}}(x)}{\xi_{\Theta|Y^{\tau}}(x)} \leq
\left\{
\begin{array}{lll}
 2\p & \text{if } Y_{\tau+1} = 1 &\text{ w.p. }\frac{1}{2},\\
 2p & \text{if } Y_{\tau+1} = 0 &\text{ w.p. }\frac{1}{2}.
\end{array}
\right.
\end{align}
Hence,  for all $x = k 2^{-i}$ where $k \in \{0,\ldots, 2^{i}-1\}$ under Case 1 we have
\begin{align}
 &\expe\left[ \left.\left(\frac{\xi_{\Theta|Y^{\tau}}(x)}{\xi_{\Theta|Y^{\tau-1}}(x) }\right)^{\lambda} \right| Y^{\tau}, N_1^{i-1} = n_1^{i-1}\right] 
 \\
&\leq 
f_1(\lambda) 
:=
\frac{(2p)^{\lambda}+(2\p)^{\lambda}}{2}.
\end{align}

\noindent
\underline{Case 2:} When median lies inside some bin we randomize the encoding.

\underline{Case 2a:} Consider $x = k 2^{-i}$ where $k \in \{0, \ldots, 2^i-1\}$ such that $\xi_{\Theta|Y^{\tau}}(x) = F_{\Theta|Y^{\tau}}(x) \leq \frac{1}{2}$. Using Lemma~\ref{lemma:bayesian_update} with probability $\pi_1^{(\tau+1)}$ the update is given as 
\begin{align}
\frac{\xi_{\Theta|Y^{\tau +1}}(x)}{\xi_{\Theta|Y^{\tau}}(x)} \leq 
\left\{
\begin{array}{ll}
\frac{p}{\frac{1}{2}+(\p-p)d_1^{(\tau)}} & \text{if } Y_{\tau+1} = 1,\\
\frac{\p}{\frac{1}{2}-(\p-p)d_1^{(\tau)}} & \text{if } Y_{\tau+1} = 0
\end{array}
\right.
\end{align}
where the probability of $\P(Y_{\tau+1} = 1\mid y_1^{\tau}) = \frac{1}{2}+(\p-p)d_1^{(\tau)}$ and $\P(Y_{\tau+1} = 0\mid y_1^{\tau}) = \frac{1}{2}-(\p-p)d_1^{(\tau)}$. Similarly with  probability $\pi_2^{(\tau+1)}$ the update is given as 
\begin{align}
\frac{\xi_{\Theta|Y^{\tau+1}}(x)}{\xi_{\Theta|Y^{\tau}}(x)} \leq 
\left\{
\begin{array}{lll}
 \frac{p}{\frac{1}{2}-(\p-p)d_2^{(\tau)}} & \text{if } Y_{\tau+1} = 1,\\
 \frac{\p}{\frac{1}{2}+(\p-p)d_2^{(\tau)}} & \text{if } Y_{\tau+1} = 0,
\end{array}
\right.
\end{align}
where $\P(Y_{\tau+1} = 1\mid y_1^{\tau}) = \frac{1}{2}-(\p-p)d_2^{(\tau)}$ and $\P(Y_{\tau+1} = 0\mid y_1^{\tau}) = \frac{1}{2}+(\p-p)d_2^{(\tau)}$.
Hence, for all $x = k 2^{-i}$ where $k \in \{0, \ldots, 2^i-1\}$ such that $\xi_{\Theta|Y^{\tau}}(x) = F_{\Theta|Y^{\tau}}(x) \leq \frac{1}{2}$, we have
\begin{align}
&\expe\left[ \left.\left(\frac{\xi_{\Theta|Y^{\tau+1}}(x)}{\xi_{\Theta|Y^{\tau}}(x) }\right)^{\lambda} \right| Y^{\tau}, N_1^{i-1} = n_1^{i-1}\right]
\\
&\leq 
f_2(\lambda)
:=
\pi_1^{(\tau+1)} g_{\lambda}\left(d_1^{(\tau)}\right)+ \pi_2^{(\tau+1)} g_{\lambda}\left(-d_2^{(\tau)}\right),
\end{align}
where we define
\begin{align}
g_{\lambda}(d)
: =
\frac{p^{\lambda}}{\left( \frac{1}{2}+(\p-p)d\right)^{\lambda - 1}}
+
\frac{\p^{\lambda}}{\left( \frac{1}{2}-(\p-p)d\right)^{\lambda - 1}}.
\end{align}

\noindent
\underline{Case 2b:} Consider $x = k 2^{-i}$ where $k \in \{0,\ldots, 2^i-1\}$ such that $\xi_{\Theta|Y^{\tau}}(x) = 1 - F_{\Theta|Y^{\tau}}(x) \leq \frac{1}{2}$. Using Lemma~\ref{lemma:bayesian_update} with probability $\pi_1^{(\tau+1)}$ the update is given as 
\begin{align}
\frac{\xi_{\Theta|Y^{\tau+1}}(x)}{\xi_{\Theta|Y^{\tau}}(x)} \leq 
\left\{
\begin{array}{ll}
\frac{\p}{\frac{1}{2}+(\p-p)d_1^{(\tau)}} & \text{if } Y_{\tau+1} = 1,\\
\frac{p}{\frac{1}{2}-(\p-p)d_1^{(\tau)}} & \text{if } Y_{\tau+1} = 0,
\end{array}
\right.
\end{align}
where $\P(Y_{\tau +1} = 1\mid y_1^{\tau}) = \frac{1}{2}+(\p-p)d_1^{(\tau)}$ and $\P(Y_{\tau +1} = 0\mid y_1^{\tau}) = \frac{1}{2}-(\p-p)d_1^{(\tau)}$. Similarly, with probability $\pi_2^{(\tau+1)}$ the update is given as 
\begin{align}
\frac{\xi_{\Theta|Y^{\tau+1}}(x)}{\xi_{\Theta|Y^{\tau}}(x)} \leq 
\left\{
\begin{array}{ll}
\frac{\p}{\frac{1}{2}-(\p-p)d_2^{(\tau)}} & \text{if } Y_{\tau} = 1,\\
\frac{p}{\frac{1}{2}+(\p-p)d_2^{(\tau)}} & \text{if } Y_{\tau} = 0,
\end{array}
\right.
\end{align}
where $\P(Y_{\tau +1} = 1\mid y_1^{\tau}) = \frac{1}{2}-(\p-p)d_1^{(\tau)}$ and $\P(Y_{\tau +1} = 0\mid y_1^{\tau}) = \frac{1}{2}+(\p-p)d_1^{(\tau)}$. Hence, $x = k 2^{-i}$ where $k \in \{0, \ldots, 2^i-1\}$ such that $\xi_{\Theta|Y^{\tau}}(x) = 1 - F_{\Theta|Y^{\tau}}(x) \leq \frac{1}{2}$, we have
\begin{align}
&\expe\left[ \left.\left(\frac{\xi_{\Theta|Y^{\tau+1}}(x)}{\xi_{\Theta|Y^{\tau}}(x) }\right)^{\lambda} \right| Y^{\tau}, N_1^{i-1} = n_1^{i-1}\right]
\\
&\leq 
f_3(\lambda)
:= 
\pi_1^{(\tau+1)} g_{\lambda}(-d_1^{(\tau)})+ \pi_2^{(\tau+1)} g_{\lambda}(d_2^{(\tau)}).
\end{align}

Now we want $\pi_1^{(\tau+1)}$ and $\pi_2^{(\tau+1)}$ which minimize the $\max\{f_2(\lambda), f_3(\lambda)\}$. Hence we choose $\pi_1^{(\tau+1)}$ and $\pi_2^{(\tau+1)}$ for which $f_2(\lambda) = f_3(\lambda)$ and obtain
\begin{align}
\pi_1^{(\tau+1)} = \frac{g_{\lambda}\left(d_2^{(\tau)}\right)-g_{\lambda}\left(-d_2^{(\tau)}\right)}{g_{\lambda}\left(d_1^{(\tau)}\right)-g_{\lambda}\left(-d_1^{(\tau)}\right)+g_{\lambda}\left(d_2^{(\tau)}\right)-g_{\lambda}\left(-d_2^{(\tau)}\right)}.
\end{align}
Note that for the above choice of $\pi_1^{(\tau)}$ and $\pi_2^{(\tau)}$ we have
\begin{align}
&\max_{d_1^{(\tau)}, d_2^{(\tau)} }\max\{f_2(\lambda), f_3(\lambda)\} 
\\
&\leq \max_{d_1^{(\tau)}, d_2^{(\tau)} } \left\{\frac{f_2(\lambda) +f_3(\lambda)}{2}\right\}
\\
&= \frac{1}{2}\pi_1^{(\tau+1)} \max_{d_1^{(\tau)} }\left\{g_{\lambda}(d_1^{(\tau)})+g_{\lambda}(-d_1^{(\tau)})\right\}
\\
& \hspace{1.2cm}+ \frac{1}{2}\pi_2^{(\tau+1)}\max_{d_2^{(\tau)}}\left\{g_{\lambda}(d_2^{(\tau)})+g_{\lambda}(-d_2^{(\tau)})\right\}
\\
&= \max_{d } \frac{1}{2} \left\{g_{\lambda}(d)+g_{\lambda}(-d)\right\}.
\end{align}
Furthermore, we have
\begin{align}
&\frac{g_{\lambda}(d)+g_{\lambda}(-d)}{2}
\\
&
=\left( \frac{(2p)^{\lambda} + (2\p)^{\lambda}}{4}\right)
\left( \left( 1+2(\p-p)d\right)^{1-\lambda}+
\right.
\\
&\hspace{2cm}\left.
+
\left( 1-2(\p-p)d\right)^{1-\lambda}\right).
\end{align}
Since $d_1, d_2 \leq \frac{1}{2}$ we have $2(\p-p)d_1, 2(\p-p)d_2 \leq 1$. If $0 < \lambda \leq 1$, then using the inequality $(1-x)^{\lambda} + (1+x)^{\lambda} \leq 2$, we have
\begin{align}
\max_{d}\{g_{\lambda}(d)+g_{\lambda}(-d)\}
\leq 
\frac{ (2p)^{\lambda} + (2\p)^{\lambda}}{2}.
\end{align} 
Therefore, for all three cases we can upper bound as follows
\begin{align}
\max_{d_1^{(\tau)}, d_2^{(\tau)}}\max\{f_1(\lambda), f_2(\lambda), f_3(\lambda)\} \leq \frac{ (2p)^{\lambda} + (2\p)^{\lambda}}{2}.
\end{align}
Hence, we have the assertion of the lemma.
\end{IEEEproof}

\begin{lemma}
\label{lemm:posterior_shape_random_arr}

For $t \geq 1$ consider $i \in \{1, \ldots, \lceil \frac{t}{n_{\min}}\rceil\}$. Consider a sample path where $N_1^{i-1} = n_1^{i-1}$ and $b(t) > i$. Then, using causal posterior matching strategy the following holds true for all $\lambda > 0$ such that $\psi(\lambda) - \frac{1}{n_{\min}}> 0$: 
\begin{align}
&\sum_{k=1}^{2^{i}-1}\expe\left[\left.\xi^{\lambda}_{\Theta | Y^{t_{i}-1}}(k 2^{-i})\right| N_1^{i-1}= n_1^{i-1}, T_i < t\right]
\\
&\leq 
\frac{1}{1-2^{-\left(\psi(\lambda)-\frac{1}{n_{\min}}\right)n_{\min}}}.
\end{align}
\end{lemma}

\begin{IEEEproof}
The first bit has been encoded for all times instants from $1$ to $t_{i}-1$, hence for $k = 2^{i-1}$ we have 
\begin{align*}
\expe\left[\left. \xi^{\lambda}_{\Theta | Y^{t_{i-1}}}(k 2^{-i})\right| N_1^{i-1} = n_1^{i-1}, T_i < t\right] 
\leq 
2^{-\psi(\lambda)(t_{i}-1)}.
\end{align*}
Note that there are $2^{i -1}$ tails which satisfy the above equation. Similarly, for $1\leq \ell \leq i-1$ the $\ell$th bit has been encoded since time $t_{\ell}$ to $t_{i}-1$ we have that there are $2^{i-\ell}$ tails of the remaining ones which are less than $2^{-\psi(\lambda)(t_{i} - t_{\ell})}$. Therefore we have 
\begin{align}
&\sum_{k=0}^{2^{i}-1}\expe\left[\left.\xi^{\lambda}_{\Theta | Y^{t_{i}-1}}(k 2^{-i})\right| N_1^{i-1} = n_1^{i-1}, T_i < t\right]
\\
&\leq 
\sum_{\ell = 0}^{i} 2^{i-\ell}2^{-\psi(\lambda)(t_{i} - t_{\ell})}
\\
&\overset{(a)}\leq 
\sum_{\ell = 0}^{i} 2^{i-\ell}2^{-\psi(\lambda)n_{\min}(i-\ell)}
\\
& = \sum_{\ell = 0}^{i}2^{-\left(\psi(\lambda)-\frac{1}{n_{\min}}\right)n_{\min}(i-\ell)}
\\
& \leq 
\sum_{\ell=0}^{\infty}2^{-\left(\psi(\lambda)-\frac{1}{n_{\min}}\right)n_{\min}\ell}
\\
&\overset{(b)}= \frac{1}{1-2^{-\left(\psi(\lambda)-\frac{1}{n_{\min}}\right)n_{\min}}},
\end{align}
where $(a)$ follows from $t_{i}-t_{\ell} \geq n_{\min}(i-\ell)$ and $(b)$ follows from that fact that $\psi(\lambda)-\frac{1}{n_{\min}} > 0$. 
\end{IEEEproof}

\begin{lemma}
\label{lemm:posterior_shape}
For $t \geq 1$ consider $i \in \{1, \ldots, \lceil \frac{t}{n_{\max}}\rceil\}$. Let $t_{i}:= n_{\max}(i-1)+1$. Then, using causal posterior matching strategy the following holds true for all $\lambda > 0$ such that $\psi(\lambda) - \frac{1}{n_{\max}}> 0$: 
\begin{align}
\sum_{k=0}^{2^{i}-1}\expe\left[\xi^{\lambda}_{\Theta | Y^{t_{i}-1}}(k 2^{-i})\right]
\leq 
\frac{1}{1-2^{-\left(\psi(\lambda)-\frac{1}{n_{\max}}\right)n_{\max}}}.
\end{align}
\end{lemma}

\begin{IEEEproof}
The first bit has been encoded for all times instants from $1$ to $t_{i}-1$, hence for $k = 2^{i-1}$ we have 
\begin{align}
\expe\left[\xi^{\lambda}_{\Theta | Y^{T_{i-1}}}(k 2^{-i})\right] 
\leq 
= 2^{-\psi(\lambda)(t_{i}-1)}.
\end{align}
Note that there are $2^{i -1}$ tails which satisfy the above equation. Similarly, for $1\leq \ell \leq i-1$ the $\ell$th bit has been encoded since time $t_{\ell}$ to $t_{i}-1$ we have that there are $2^{i-\ell}$ tails of the remaining ones which are less than $2^{-\psi(\lambda)(t_{i} - t_{\ell})}$. Noting that $t_{i}-t_{\ell} = n_{\max}(i-\ell)$ and following the proof of Lemma~\ref{lemm:posterior_shape_random_arr} we have the assertion of the lemma.
\end{IEEEproof}

\bibliographystyle{IEEEtran}
\bibliography{toly}

\end{document}